\documentclass[prl,twocolumn,showkeys,amsmath,amssymb]{revtex4-1}
\usepackage{graphicx}
\usepackage{amsmath}
\usepackage{bm}
\usepackage{yhmath}
\usepackage{hyperref}
\usepackage{subfigure}
\usepackage{color}
\usepackage{cases}
\usepackage{subfigure}
\usepackage{dcolumn,booktabs,bm}
\usepackage{amsfonts,amssymb,stmaryrd,latexsym,amsmath}
\usepackage{textcomp}

\usepackage{array}
\newcounter{RomanNumber}

\newcommand{\lyxmathsym}[1]{\ifmmode\begingroup\def\b@ld{bold}
  \text{\ifx\math@version\b@ld\bfseries\fi#1}\endgroup\else#1\fi}

\UseRawInputEncoding
\begin{document}

\title{Ab initio calculation of molecular pentaquark magnetic moments in heavy pentaquark chiral perturbation theory}
\author{Hao-Song Li}\email{haosongli@nwu.edu.cn}\affiliation{School of Physics, Northwest University, Xian 710127, China}\affiliation{Shaanxi Key Laboratory for Theoretical Physics Frontiers, Xi¡¯an 710127, China}\affiliation{
Peng Huanwu Center for Fundamental Theory, Xi¡¯an 710127, China}

\begin{abstract}
We propose a new heavy pentaquark chiral perturbation theory for the recently observed hidden-charm pentaquark states by LHCb Collaboration. With the ab initio constructed chiral Lagrangians, we present a parameter-free calculation of the octet molecular pentaquark magnetic moments up to one-loop level. We improve the quark model description of the data when we include the leading SU(3) breaking effects coming from the one-loop corrections. Without any experimental inputs, our predictions are so simple and unique that we regard them as a theoretical benchmark to be compared with experiments as well as other theoretical models.
\end{abstract}

\maketitle

The hidden-charm pentaquark state was firstly observed by LHCb collaboration in 2015~\cite{LHCb:2015yax}, which
appear to be inconsistent with the predictions of the conventional quark model~\cite{Gell-Mann:1964ewy}. Since then, experimental collaborations have shown great interest to the pentaquark states, an increasing amount of data has produced many pentaquark states. In 2019, $P^{N}_{\psi}(4312)^{^{+}}$, $P^{N}_{\psi}(4440)^{^{+}}$  and $P^{N}_{\psi}(4457)^{^{+}}$ were reported in the updated analyses of the LHCb collaboration~\cite{LHCb:2019kea}. In 2020, the LHCb Collaboration observed the hidden-charm strange pentaquark state $P_{\psi{}s}^{\Lambda}(4459)^{0}$ through an amplitude analysis of the $\Xi^{-}_b\to{}J/\psi{}\Lambda{}K^{-}$ decay \cite{LHCb:2020jpq}. In 2021,  $P^{N}_{\psi}(4337)^{^{+}}  $  was observed by the LHCb Collaboration in the $B_{s}^{0}\to J/\psi p\bar{p}$ decay~\cite{LHCb:2021chn}. Recently, the LHCb collaboration observed a new structure $P_{\psi{}s}^{\Lambda}(4338)^{0}$ in the $B^{-}\to{}J/\psi{}\Lambda\bar{p}$ decay~\cite{LHCb:2022ogu}.

With the discovery of so many hidden-charm pentaquark states in the experiments, theorists have made intense speculations on the nature of these states~\cite{Deng:2022vkv,Wang:2019nvm,Xiao:2019gjd,Liu:2020hcv,Peng:2020hql,Zhu:2021lhd,Du:2021bgb,
Chen:2021tip,Yang:2021pio,Ozdem:2021btf}. Due to their proximity to various charmed meson-baryon thresholds, those candidates of the hidden-charm pentaquark states can be explained as the molecular states. It is reasonable to speculate that the full molecular multiplet of the hidden-charm molecular pentaquarks will be observed in the coming future.

The hidden-charm molecular pentaquark states are composed of the corresponding singly charmed baryons and anti-charmed mesons. Considering the direct product $3\otimes3\otimes3 = 1\oplus8_1\oplus8_2\oplus10$ and that the $P^{N}_{\psi}$ pentaquark states are in the isospin $I=1/2$ configuration~\cite{Chen:2015loa,Chen:2015moa}, we construct the chiral lagrangians for octet hidden-charm pentaquark states of $SU(3)$ flavor symmetry in this work.

To determine the nature of the pentaquark states, it is crucial to understand their internal structures. The hadron magnetic moments provide valuable insight in describing the inner structures of hadrons and understanding the mechanism of strong interactions at low-energy. We first investigated the magnetic moments of the pentaquark state with quark model in Ref.~\cite{Gao:2021hmv,Guo:2023fih}, however it is difficult to include the Goldstone boson cloud effect.

In fact, chiral perturbation theory (ChPT)~\cite{Weinberg:1978kz} is quite helpful to analyze the low-energy interactions. To consider the chiral corrections of pentaquark we develop the heavy pentaquark chiral perturbation theory (HPChPT) similar to heavy baryon chiral perturbation theory (HBChPT)~\cite{Jenkins:1990jv,Jenkins:1992pi,Bernard:1992qa,Bernard:1995dp} which offers a useful power counting and ease of calculation of Feynman diagrams. In the heavy pentaquark limit, the pentaquark field $B$ can be decomposed into the large component $P$ and the small component $L$
\begin{eqnarray}
B=e^{-iM_{P}v\cdot x}(P+L),\nonumber
\end{eqnarray}
\begin{eqnarray}
P=e^{iM_{P}v\cdot x}\frac{1+v\hspace{-0.5em}/}{2}B,~
L=e^{iM_{P}v\cdot x}\frac{1-v\hspace{-0.5em}/}{2}B,\nonumber
\end{eqnarray}
where $M_{P}$ is the mass of pentaquark state and $v_{\mu}=(1,\vec{0})$ is the velocity of the pentaquark state. The octet hidden-charm molecular pentaquark $P_n$ reads
		\begin{equation}
			\setlength{\arraycolsep}{0.1pt}
			P_n=
			\left(		
			\begin{array}{ccc}
				\frac{1}{\sqrt{2}}{{}P_{\psi s}^{\Sigma}}^{0}+\frac{1}{\sqrt{6}}{{}P_{\psi s}^{\Lambda}}^{0}
				&{{}P_{\psi s}^{\Sigma}}^{+}
				&{{}P_{\psi}^{N}}^{+}
				\\
				{{}P_{\psi s}^{\Sigma}}^{-}
				&-\frac{1}{\sqrt{2}}{{}P_{\psi s}^{\Sigma}}^{0}+\frac{1}{\sqrt{6}}{{}P_{\psi s}^{\Lambda}}^{0}
				&{{}P_{\psi}^{N}}^{0}
				\\
				{{}P_{\psi ss}^{N}}^{-}
				&{{}P_{\psi ss}^{N}}^{+}
				&\frac{2}{\sqrt{3}}{{}P_{\psi s}^{\Lambda}}^{0}
				\nonumber\\
			\end{array}
			\right).
		\end{equation}
The flavor wave functions of the octet hidden-charm molecular pentaquark under $SU(3)$ symmetry are shown in Table~\ref{tab111} (details in Ref.~\cite{{Li111}}).
	\begin{table}[htbp]
			\centering
			\caption{$8_{1}$ and $8_{2}$ flavor wave functions}
			\label{tab111} 		
		\begin{tabular}{c|c|c}
				\toprule[1.0pt]
				\toprule[1.0pt]
				States &  Flavor & Wave functions
				\\
				\hline
					{${P_{\psi}^{N+}}$}
				&	$8_{1}$
				&	$-\sqrt{\frac{1}{3}}\Sigma_c^{+}\bar{D}^{(*)0}+\sqrt{\frac{2}{3}}\Sigma_{c}^{++}D^{(*)-}$
				\\

				&	$8_{2}$
				&	$\Lambda_{c}^{+}\Bar{D}^{(*)0}$
				\\
				\hline

					{${P_{\psi}^{N0}}$}
				&	$8_{1}$
				&	$\sqrt{\frac{1}{3}}\Sigma_c^{+}D^{(*)-}-\sqrt{\frac{2}{3}}\Sigma_{c}^{0}\bar{D}^{(*)0}$
				\\

				&	$8_{2}$
				&	$\Lambda_{c}^{+}D^{(*)-}$
				\\
				\hline

					{${P_{\psi s}^{\Sigma+}}$}
				&	$8_{1}$
				&	$\sqrt{\frac{1}{3}}\Xi_c^{\prime+}\bar{D}^{(*)0}-\sqrt{\frac{2}{3}}\Sigma_{c}^{++}D_s^{(*)-}$
				\\

				&	$8_{2}$
				&	$\Xi_{c}^{+}\Bar{D}^{(*)0}$
				\\
				\hline

					{${P_{\psi s}^{\Sigma0}}$}
				&	$8_{1}$
				&	$\sqrt{\frac{1}{6}}\Xi_c^{\prime+}D^{(*)-}+\sqrt{\frac{1}{6}}\Xi_{c}^{\prime0}\bar{D}^{(*)0}-\sqrt{\frac{2}{3}}\Sigma_{c}^{+}D_s^{(*)-}$
				\\

				&	$8_{2}$
				&	$\sqrt{\frac{1}{2}}\Xi_c^{+}D^{(*)-} + \sqrt{\frac{1}{2}}\Xi_{c}^{0}\bar{D}^{(*)0} $
				\\
				\hline

					{${P_{\psi s}^{\Lambda0}}$}
				&	$8_{1}$
				&	$\sqrt{\frac{1}{2}}\Xi_c^{\prime+}D^{(*)-} - \sqrt{\frac{1}{2}}\Xi_{c}^{\prime0}\bar{D}^{(*)0} $
				
				\\

				&	$8_{2}$
				&	$\sqrt{\frac{1}{6}}\Xi_c^{+}D^{(*)-}-\sqrt{\frac{1}{6}}\Xi_{c}^{0}\bar{D}^{(*)0}-\sqrt{\frac{2}{3}}\Lambda_{c}^{+}D_s^{(*)-}$
				\\
				\hline

					{${P_{\psi s}^{\Sigma-}}$}
				&	$8_{1}$
				&	$\sqrt{\frac{1}{3}}\Xi_c^{\prime0}D^{(*)-}-\sqrt{\frac{2}{3}}\Sigma_{c}^{0}D_s^{(*)-}$
				\\

				&	$8_{2}$
				&	$\Xi_{c}^{0}D^{(*)-}$
				\\
				\hline

					{${P_{\psi ss}^{N0}}$}
				&	$8_{1}$
				&	$\sqrt{\frac{1}{3}}\Xi_c^{\prime+}D_{s}^{(*)-}-\sqrt{\frac{2}{3}}\Omega_{c}^{0}\bar{D}^{(*)0}$
				\\
				
				&	$8_{2}$
				&	$\Xi_{c}^{+}D_{s}^{(*)-}$
				\\
				\hline
	{${P_{\psi ss}^{N-}}$}
				&	$8_{1}$
				&	$\sqrt{\frac{1}{3}}\Xi_c^{\prime0}D_{s}^{(*)-}-\sqrt{\frac{2}{3}}\Omega_{c}^{0}D^{(*)-}$
				\\

				&	$8_{2}$
				&	$\Xi_{c}^{0}D_{s}^{(*)-}$
				\\
			\toprule[1.0pt]
			\toprule[1.0pt]
			\end{tabular}	
		\end{table}

In this work, we investigate the magnetic moments of spin-$\frac{1}{2}$ octet hidden-charm molecular pentaquark family in HPChPT. We explicitly consider the molecular pentaquark intermediate states in the loop calculation. We use quark model to determine the corresponding low energy constants (LECs) and calculate the magnetic moments order by order. We follow the standard power counting scheme, the chiral
order $D_{\chi}$ of a given diagram is given by~\cite{Ecker:1994gg}
\begin{equation}
D_{\chi}=4N_{L}-2I_{M}-I_{P}+\sum_{n}nN_{n}, \label{Eq:Power
counting}
\end{equation}
where $N_{L}$ is the number of loops, $I_{M}$ is the number of
internal pion lines, $I_{P}$ is the number of internal molecular pentaquark lines and $N_{n}$ is the number of the vertices from the $n$th order Lagrangians.

To calculate the chiral corrections to the magnetic moment of spin-$\frac{1}{2}$ pentaquark state, we follow the
 basic definitions of the pseudoscalar mesons and the spin-$\frac{1}{2}$ baryon chiral effective Lagrangians in Refs.~\cite{Bernard:1995dp,Li:2017cfz} to construct the relevant chiral Lagrangians.
The pseudoscalar meson fields read
\begin{equation}
\phi=\left(\begin{array}{ccc}
\pi^{0}+\frac{1}{\sqrt{3}}\eta & \sqrt{2}\pi^{+} & \sqrt{2}K^{+}\\
\sqrt{2}\pi^{-} & -\pi^{0}+\frac{1}{\sqrt{3}}\eta & \sqrt{2}K^{0}\\
\sqrt{2}K^{-} & \sqrt{2}\bar{K}^{0} & -\frac{2}{\sqrt{3}}\eta
\end{array}\right).\nonumber
\end{equation}
 The chiral connection and axial vector
field read \cite{Bernard:1995dp,Scherer:2002tk}:
\begin{eqnarray}
\Gamma_{\mu}&=&\frac{1}{2}\left[u^{\dagger}(\partial_{\mu}-ir_{\mu})u+u(\partial_{\mu}-il_{\mu})u^{\dagger}\right],
\\u_{\mu}&=&\frac{1}{2}i\left[u^{\dagger}(\partial_{\mu}-ir_{\mu})u-u(\partial_{\mu}-il_{\mu})u^{\dagger}\right],
\end{eqnarray}
where $u^{2}=\mathit{U}=\exp(i\phi/f_{0})$, $r_{\mu}=l_{\mu}=-eQA_{\mu}$, $Q=\rm{diag}(2/3,-1/3,-1/3)$. We use the pseudoscalar meson decay constants
$f_{\pi}\approx$ 92.4 MeV, $f_{K}\approx$ 113 MeV and
$f_{\eta}\approx$ 116 MeV.

The octet hidden-charm molecular pentaquark states have three spin configurations $J^{P}(J_{b} ^{P_{b}}\otimes J_{m} ^{P_{m}})$: $\frac{1}{2}^{-}(\frac{1}{2}^{+}\otimes0^{-})$, $\frac{1}{2}^{-}(\frac{1}{2}^{+}\otimes1^{-})$, and $\frac{3}{2}^{-}(\frac{1}{2}^{+}\otimes1^{-})$, $J_{b} ^{P_{b}}$ and $J_{m} ^{P_{m}}$ correspond to the angular momentum and parity of baryon and meson, respectively. We only consider the $J^{P}=\frac{1}{2}^{-}(\frac{1}{2}^{+}\otimes0^{-})$  pentaquark state in this work.

The free Lagrangian of pentaquark state reads
\begin{equation}
\mathcal{L}_{0}^{(1)}={\rm Tr}[\bar{P}(iv\cdot
D)P], \label{Eq:baryon1}
\end{equation}
\begin{eqnarray}
D_{\mu}P&=&\partial_{\mu}P+\Gamma_{\mu}P.
\end{eqnarray}
The interaction Lagrangians of $8_{1}$ and $8_{2}$ flavor pentaquark states read
\begin{equation}
\mathcal{L}_{\rm
int1}=2g_{1}{\rm Tr}(\bar{P}_1 S_{\mu}\{u^{\mu},P_1\}) +2f_{1}{\rm Tr}(\bar{P}_1S_{\mu}[u^{\mu},P_1]),
\label{Eq:baryon21}
\end{equation}
\begin{equation}
\mathcal{L}_{\rm
int2}=2g_{4}{\rm Tr}(\bar{P}_2 S_{\mu}\{u^{\mu},P_2\}) +2f_{4}{\rm Tr}(\bar{P}_2S_{\mu}[u^{\mu},P_2]),
\label{Eq:baryon22}
\end{equation}
where $P_{1,2}$ represent the $8_{1,2}$ flavor pentaquark states, $S_{\mu}$ is the
covariant spin-operator. The $\phi PP$ coupling constants $f_{1,4}$ and $g_{1,4}$ are estimated in our another work in Ref.~\cite{{Li111}}. For the pseudoscalar mesons  masses, we use $m_{\pi}=0.140$ GeV, $m_{K}=0.494$ GeV, and
$m_{\eta}=0.550$ GeV.

We also need the lowest order $\mathcal{O}(p^{2})$ Lagrangian contributes to the
magnetic moments of the $8_{1,2}$ flavor pentaquark states at the tree level
\begin{eqnarray}
\mathcal{L}_{\mu1}^{(2)}&=&b_{p1}{\rm Tr}(\frac{-i}{4M_{B}}\bar{P}_{1}[S^{\mu},S^{\nu}]\{F_{\mu\nu}^{+},P_{1}\})\nonumber\\
&&
+b_{p2}{\rm Tr}(\frac{-i}{4M_{B}}\bar{P}_{1}[S^{\mu},S^{\nu}][F_{\mu\nu}^{+},P_{1}]),\\
\mathcal{L}_{\mu2}^{(2)}&=&b_{p3}{\rm Tr}(\frac{-i}{4M_{B}}\bar{P}_{2}[S^{\mu},S^{\nu}]\{F_{\mu\nu}^{+},P_{2}\})\nonumber\\
&&
+b_{p4}{\rm Tr}(\frac{-i}{4M_{B}}\bar{P}_{2}[S^{\mu},S^{\nu}][F_{\mu\nu}^{+},P_{2}]).
\label{Eq:MM1}
\end{eqnarray}
The coefficients $b_{p(1, 2, 3, 4)}$ are LECs.
The chirally covariant QED field strength tensor $F_{\mu\nu}^{\pm}$
is defined as
\begin{eqnarray} \nonumber
F_{\mu\nu}^{\pm} & = & u^{\dagger}F_{\mu\nu}^{R}u\pm
uF_{\mu\nu}^{L}u^{\dagger},\\
F_{\mu\nu}^{R} & = &
\partial_{\mu}r_{\nu}-\partial_{\nu}r_{\mu}-i[r_{\mu},r_{\nu}],\nonumber\\
F_{\mu\nu}^{L} & = &
\partial_{\mu}l_{\nu}-\partial_{\nu}l_{\mu}-i[l_{\mu},l_{\nu}],\nonumber
\end{eqnarray}
We use the nucleon mass $M_B=0.938$ GeV, so that we can express the magnetic moments directly in nuclear magnetons.

%

\begin{figure}[tbh]
\centering
\includegraphics[width=0.9\hsize]{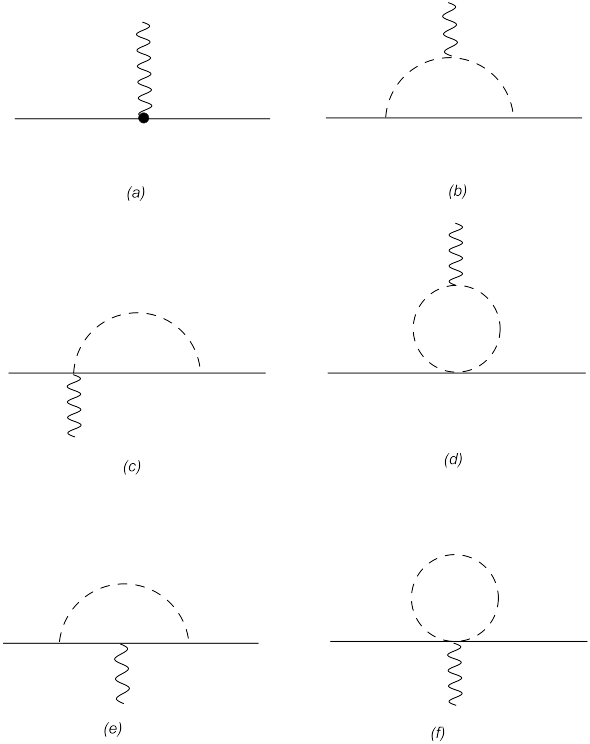}
\caption{Feynman diagrams where octet pentaquark state is
denoted by the solid line. The dot represent
$\mathcal{O}(p^{2})$ coupling. The dashed and wiggly
lines represent the pseudoscalar meson and photon
respectively.}\label{fig:allloop}

\end{figure}
\begin{table*}
  \centering
  \caption{The coefficients of the loop corrections to the
pentaquark magnetic moments from Figs.
\ref{fig:allloop}(b).} \label{table:ab}
\begin{tabular}{c|cccccccc}
\toprule[1pt]\toprule[1pt]
 States & $P_{\psi}^{N+}$ & $P_{\psi}^{N0}$ & $P_{\psi s}^{\Sigma+}$
 & $\ \ \ P_{\psi s}^{\Sigma0}\ \ \ $ & $\ \ \ P_{\psi s}^{\Lambda0}\ \  \ $ & $P_{\psi s}^{\Sigma-}$  & $P_{\psi ss}^{N0}$ &$P_{\psi ss}^{N-}$\tabularnewline
\midrule[1pt]
$\zeta_1=\zeta_2$ & $\frac{1}{3}$ & $-\frac{2}{3}$ & $\frac{1}{3}$ & $\frac{1}{3}$ & $-\frac{1}{3}$
&  $\frac{1}{3}$&  $-\frac{2}{3}$ &$\frac{1}{3}$\tabularnewline
\hline
$\kappa_1=\kappa_2$ & 1 &  0 & 1 & 0
& 0 & -1& 0& -1\tabularnewline
\midrule[1pt]

$\beta_1^{\pi}$ & $2 (f_1+g_1)^2$ & $-2 (f_1+g_1)^2$ & $4f_1^2+\frac{4}{3}g_1^2$ & $0$  & $0$
&  $-4 f_1^2-\frac{4}{3}g_1^2$&  $2 (g_1-f_1)^2$ &$-2 (g_1-f_1)^2$\tabularnewline
\hline
$\beta_1^{K}$ & $4 f_1^2+\frac{4}{3}g_1^2$ &  $2 (g_1-f_1)^2$ & $2 (f_1+g_1)^2$ & $4f_1g_1$ & $-4f_1g_1$
& $-2 (g_1-f_1)^2$ & $-2 (f_1+g_1)^2$& $-2 (f_1+g_1)^2$\tabularnewline
\midrule[1pt]
$\beta_2^{\pi}$ & $2 (f_4+g_4)^2$ & $-2 (f_4+g_4)^2$ & $4f_4^2+\frac{4}{3}g_4^2$ & $0$  & $0$
&  $-4 f_4^2-\frac{4}{3}g_4^2$&  $2 (g_4-f_4)^2$ &$-2 (g_4-f_4)^2$\tabularnewline
\hline
$\beta_2^{K}$ & $4 f_4^2+\frac{4}{3}g_4^2$ &  $2 (g_4-f_4)^2$ & $2 (f_4+g_4)^2$ & $4f_4g_4$ & $-4f_4g_4$
& $-2 (g_4-f_4)^2$ & $-2 (f_4+g_4)^2$& $-2 (f_4+g_4)^2$\tabularnewline
\bottomrule[1pt]\bottomrule[1pt]
\end{tabular}

\end{table*}
There are six Feynman diagrams contributing to the pentaquark magnetic moments to one-loop
order shown in Fig.~\ref{fig:allloop}. The tree-level Lagrangians in Eq.~(\ref{Eq:MM1}) contribute to $8_{1,2}$ flavor pentaquark magnetic moments at leading order (LO) as shown in diagram (a)
\begin{eqnarray}
\mu_{P1}^{(\rm tree)}& =&\zeta_1b_{p1}+\kappa_1b_{p2},\\
\label{eq:mutree1}
\mu_{P2}^{(\rm tree)}& =&\zeta_2b_{p3}+\kappa_2b_{p4},
\label{eq:mutree2}
\end{eqnarray}
where $\mu_{P(1,2)}^{(\rm tree)}$ represent the LO $8_{1,2}$ flavor pentaquark magnetic moments, the coefficients $\zeta_{1,2}$ and $\kappa_{1,2}$ for $8_{1,2}$ flavor pentaquark are listed in Table~\ref{table:ab}. Due to the SU(3)-symmetry, the LO contributions show the Coleman-Glashow relations similar to the nucleon octet.

 The diagram (c) vanishes in the heavy pentaquark mass limit. The diagrams (d), (e) and (f) contribute to the pentaquark magnetic moments to next-to-next-to-leading order. Thus, only diagrams (b) account for the next-to-leading order (NLO) corrections which are induced by the corresponding breaking in the pseudoscalar meson masses. The NLO corrections to the pentaquark magnetic moments can be expressed as

\begin{eqnarray}
\mu_{P(1, 2)}^{(\rm loop)}& = &\sum_{\phi=\pi,K}\frac{-m_{\phi}M_{B}\beta_{(1, 2)}^{\phi}}{64\pi f_{\phi}^2}
\label{eq:mu2Loop}
\end{eqnarray}
$\mu_{P(1, 2)}^{(\rm loop)}$ represent the loop corrections to the $8_{1,2}$ flavor pentaquark magnetic moments. We collect the coefficients
$\beta_{1,2}^\phi$  explicit expressions in Table~\ref{table:ab}.

In ChPT, the LECs should be fitted through experimental inputs. As there are not any experimental data of the pentaquark magnetic moments so far, in this work,
we use quark model to determine the LO tree level magnetic moment. The magnetic moments in the quark model are the matrix elements of the following operator,
\begin{equation}
\vec{\mu}=\sum_i\mu_i\vec{\sigma}^i, \label{magmomen}\nonumber
\end{equation}
where $\mu_i$ is the magnetic moment of the quark:
\begin{equation}
\mu_i={e_i\over 2m_i},\quad i=u,d,s,c.\nonumber
\end{equation}

With the wave functions of the octet hidden-charm molecular pentaquark states in Table~\ref{tab111}  (details in Ref.~\cite{{Li111}}),
 we give the LO pentaquark magnetic moments in the second and third columns in Table~\ref{Magnetic moments81}. When considering SU(3) flavor symmetry ($m_{u}=m_{d}=m_{s}$) and heavy quark symmetry ($m_{c}=\infty$), we obtain the LECs $b_{p1}=0.93$, $b_{p2}=1.55$, $b_{p3}=b_{p4}=0$.
When we take the constituent quark masses as input:
$m_u = m_d = 0.336$ GeV, $m_s = 0.540$ GeV, $m_c = 1.660$ GeV, we obtain the LO pentaquark magnetic moments shown in the second row in Table~\ref{Magnetic moments}. Thus, up to NLO, there exist only four LECs: $f_{1,4}$, $g_{1,4}$
at this order.

\begin{table}[htbp]
	\caption{The $8_{1}$ and $8_{2}$ flavor pentaquark magnetic moments at tree level (in unit of $\mu_{N}$).}	\label{Magnetic moments81}
\centering	
	\begin{tabular}{c|c|c}
				\toprule[1.0pt]
				\toprule[1.0pt]
				States & $8_{1}$ LO  & $8_{2}$ LO
				\\
				\hline
				$P_{\psi}^{N+}$
				&	 $\frac{2}{9}(5\mu_u+\mu_d)-\frac{1}{3}\mu_c$
				&	$\mu_c$
				\\
				\hline

					$P_{\psi}^{N0}$
				&	$\frac{2}{9}(\mu_u+5\mu_d)-\frac{1}{3}\mu_c$
				&	$\mu_c$
				\\
                \hline

					$P_{\psi s}^{\Sigma+}$
				&	$\frac{2}{9}(5\mu_u+\mu_s)-\frac{1}{3}\mu_c$
				&	$\mu_c$
			
				\\
				\hline

					$P_{\psi s}^{\Sigma0}$
				&	$\frac{1}{9}(5\mu_u+5\mu_d+2\mu_s)-\frac{1}{3}\mu_c$
				&	$\mu_c$			\\
				\hline

					$P_{\psi s}^{\Lambda0}$
				&	$\frac{1}{3}(\mu_u+\mu_d+2\mu_s)-\frac{1}{3}\mu_c$
				&	$\mu_c$
				
				\\
				\hline

				$P_{\psi s}^{\Sigma-}$
				&	$\frac{2}{9}(5\mu_d+\mu_s)-\frac{1}{3}\mu_c$
				&	$\mu_c$				\\
				\hline

					$P_{\psi ss}^{N0}$
				&	$\frac{2}{9}(\mu_u+5\mu_s)-\frac{1}{3}\mu_c$
				&	$\mu_c$				\\
				
				\hline

					$P_{\psi ss}^{N-}$
				&	$\frac{2}{9}(\mu_d+5\mu_s)-\frac{1}{3}\mu_c$
				&	$\mu_c$				\\
			\toprule[1.0pt]
			\toprule[1.0pt]

			\end{tabular}

		\end{table}
To determine the coupling constants $g_{1}$ and $f_{1}$, similar to the procedure employed for the nucleon, we consider the $\pi_{0}$ meson decay of $P_{\psi}^{N^{+}}$ and ${P_{\psi ss}^{N^{0}}}$. The Lagrangians for the $\pi_{0}$ decay of ${{}P_{\psi}^{N^{+}}}$ and ${P_{\psi ss}^{N^{0}}}$ with the spin configuration $J^{P}=\frac{1}{2}^{-}(\frac{1}{2}^{+}\otimes0^{-})$ at the hadron level read
		\begin{eqnarray}
			\mathcal{L}_{P_{\psi}^{N+}}
			&=&
			\frac{1}{2}(g_{1}+f_{1}) {{}\bar{P}_{\psi}^{N+}} \gamma^{\mu} \gamma_5  \partial_{\mu}\Phi{{}P_{\psi}^{N+}},\nonumber\\
			\mathcal{L}_{{P_{\psi ss}^{N0}}}
			&=&
			\frac{1}{2}(f_{1}-g_{1}) {{}\bar{P}_{\psi ss}^{N0}} \gamma^{\mu} \gamma_5 \partial_{\mu}\Phi{{}P_{\psi  ss}^{N0}}.\nonumber
		\end{eqnarray}
At the hadron level, the $\pi_0$ hadron interaction read
		\begin{eqnarray} 	
			 \langle
			{P_{\psi}^{N+}}; \pi_{0}
			|
			\frac{g_{1}+f_{1}}{f_\pi}{{}\bar{P}_{\psi}^{N+}}\partial_{z}  {\pi_{0}}{{}P_{\psi}^{N+}}
			|
			{P_{\psi}^{N+}}
			\rangle \sim
			\frac{g_{1}+f_{1}}{2}\frac{q_z}{f_\pi},	\nonumber	\\	
			\langle
			{P_{\psi ss}^{N0}}; \pi_{0}
			|
			\frac{f_{1}-g_{1}}{f_\pi}{\bar{P}_{\psi ss}^{N0}}\partial_{z} {\pi_{0}}{P_{\psi  ss}^{N0}}
			|
			{P_{\psi ss}^{N0}}
			\rangle \sim
			\frac{f_{1}-g_{1}}{2}\frac{q_z}{f_\pi}.\nonumber
		\end{eqnarray}
At the quark level, the $\pi_0$ quark interaction read
		\begin{eqnarray}
			\langle {P_{\psi}^{N+}} , +\frac{1}{2} ;~\pi_0| \mathcal{L}_{quark}|{{}P_{\psi}^{N+}},
			+\frac{1}{2}\rangle
			\sim\frac{4}{9}\frac{q_z}{f_\pi}g_q,\nonumber
			\\
			\langle {P_{\psi ss}^{N0}} , +\frac{1}{2} ;~\pi_0| \mathcal{L}_{quark}|{P_{\psi ss}^{N0}},
			+\frac{1}{2}\rangle
			\sim\frac{1}{9} \frac{q_z}{f_\pi}g_q.\nonumber
		\end{eqnarray}\nonumber
	Compare with the axial charge of the nucleon $g_{A}$,
		\begin{eqnarray}
			\frac{\frac{1}{2}g_{A}}{\frac{5}{6}g_{q}}	
			=
			\frac{\frac{g_{1}+f_{1}}{2}}{\frac{4}{9}g_{q}}
			=
			\frac{\frac{f_{1}-g_{1}}{2}}{\frac{1}{9}g_{q}}. \nonumber
		\end{eqnarray}

\begin{table*}
\caption{The  pentaquark magnetic moments to NLO (in
unit of $\mu_{N}$).} \label{Magnetic moments}
  \centering
  \begin{tabular}{c|cccccccc}
\toprule[1pt]\toprule[1pt]
States & $\ \ \ P_{\psi}^{N+}\ \ \ $ & $\ \ \ P_{\psi}^{N0}\ \ \ $ & $\ \ \ P_{\psi s}^{\Sigma+}\ \ \ $
 & $\ \ \ P_{\psi s}^{\Sigma0}\ \ \ $ & $\ \ \ P_{\psi s}^{\Lambda0}\ \ \ $ & $\ \ \ P_{\psi s}^{\Sigma-}\ \ \ $  & $\ \ \ P_{\psi ss}^{N0}\ \ \ $ &$\ \ \ P_{\psi ss}^{N-}\ \ \ $\tabularnewline
\midrule[1pt]
$8_{1}$ LO & 1.74 & -0.75 & 1.81& 0.26 & -0.20
&  -1.29& -0.36 &-0.98\tabularnewline
\hline
$8_{1}$ NLO & -0.21&  0.06 & -0.22 & -0.07 & 0.07
& 0.07 & 0.16& 0.15\tabularnewline
\hline
$8_{1}$ Total & 1.53&  -0.69 &1.59 & 0.19 & -0.13
& -1.22 & -0.20& 0.83\tabularnewline
\midrule[1pt]
$8_{2}$ LO & 0.38 & 0.38 &0.38 &0.38 &0.38 &0.38 &0.38 &0.38 \tabularnewline
\hline
$8_{2}$ NLO & 0 &  0 & 0 & 0
& 0 & 0& 0& 0\tabularnewline
\hline
$8_{2}$ Total & 0.38 & 0.38 &0.38 &0.38 &0.38 &0.38 &0.38 &0.38 \tabularnewline

\bottomrule[1pt]\bottomrule[1pt]
\end{tabular}
\end{table*}

We obtain $f_{1} = \frac{1}{3}g_{A}=0.42$ and $g_{1} = \frac{1}{5}g_{A}=0.25$. Similarly, we can also obtain $f_{4} =g_{4}=0$. With $f_{1,4}$ and $g_{1,4}$, we obtain the chiral loop corrections to the pentaquark magnetic moments. Thus we
obtain the numerical results of the NLO pentaquark magnetic moments in Table~\ref{Magnetic moments}.

For $8_{2}$ flavor pentaquark states, it is interesting to notice that the NLO contributions to the magnetic moments are all zero due to $f_{4} =g_{4}=0$ in quark model. Thus, the $8_{2}$ flavor pentaquark magnetic moments are all equal to $\mu_{c}=0.38\mu_{N}$. At present, the magnetic moments of the recently observed pentaquark states have not been measured in the experiment, our calculation shows that if the magnetic moment of the pentaquark equal to $\mu_{c}=0.38\mu_{N}$, the pentaquark state belong to  $J^{P}=\frac{1}{2}^{-}(\frac{1}{2}^{+}\otimes0^{-})$ $8_{2}$ flavor pentaquark states whose wave functions are shown in  Table~\ref{tab111}. We are looking forward to further progresses in experiment.

For $8_{1}$ flavor pentaquark states, in order to study the convergence of the chiral expansion,
we separate the LO from NLO contributions for the $8_{1}$ flavor pentaquark magnetic moments (in unit of $\mu_{N}$), we also compare the $8_{1}$ flavor pentaquark magnetic moments with the corresponding baryon-octet magnetic moments in PDG \cite{ParticleDataGroup:2022pth}
\begin{eqnarray}
\mu_{P1_{\psi}^{N+}}&=&1.74(1-0.12)=1.53=0.55\mu_{p},\nonumber\\
\mu_{P1_{\psi}^{N0}}&=&-0.75(1-0.08)=-0.69=0.36\mu_{n},\nonumber\\
\mu_{P1_{\psi s}^{\Sigma+}}&=&1.81(1-0.12)=1.59=0.65\mu_{\Sigma^+},\nonumber\\
\mu_{P1_{\psi s}^{\Sigma0}}&=&0.26(1-0.27)=0.19,\nonumber\\
\mu_{P1_{\psi s}^{\Lambda0}}&=&-0.20(1-0.35)=-0.13=0.21\mu_{\Lambda},\nonumber\\
\mu_{P1_{\psi s}^{\Sigma-}}&=&-1.29(1-0.05)=-1.22=1.05\mu_{\Sigma^-},\nonumber\\
\mu_{P1_{\psi ss}^{N0}}&=&-0.36(1-0.44)=-0.20=0.16\mu_{\Xi^0},\nonumber\\
\mu_{P1_{\psi ss}^{N-}}&=&-0.98(1-0.15)=-0.83=1.27\mu_{\Xi^-}.\nonumber
\end{eqnarray}

It is obvious that the convergence of the chiral expansion is quite good. The reason is accessible, since the axial charges of $8_{1}$ flavor pentaquark are generally lower compared to that of the nucleon ($f_{1} = \frac{1}{3}g_{A}$, $g_{1} = \frac{1}{5}g_{A}$), thus the NLO contributions for the $8_{1}$ flavor pentaquark magnetic moments which are proportional to $g_{A}^2$ are greatly suppressed compared to the NLO contributions for the baryon-octet magnetic moments. The $8_{1}$ flavor pentaquark magnetic moments and the baryon-octet magnetic moments present a certain degree of positive correlation.

 In summary, we have investigated the
magnetic moments for the octet hidden-charm
molecular pentaquark family to NLO in the framework of HPChPT. We construct the chiral lagrangians of the octet pentaquark states and determine all free parameters
needed for the low energy dynamics of octet pentaquark with the quark model. Thus, we report on the first global study of octet molecular pentaquark magnetic moments up to one-loop
order.

Our calculations indicate that for $8_{1}$ flavor pentaquark magnetic moments, the convergence of the chiral expansion is pretty good, for $8_{2}$ flavor pentaquark states, the pentaquark magnetic moments are all equal to $\mu_{c}=0.38\mu_{N}$ even to NLO. Moreover, these calculations depend only on the spin-flavor wave functions of the pentaquark states. Without any
experimental inputs, our predictions are so simple and unique that we regard them as a theoretical benchmark to be compared with experiments as well as other theoretical models. We are looking forward to further progresses and hope our results may be useful for future
experimental measurement.

\section*{ACKNOWLEDGMENTS}

	This project is supported by the National Natural Science Foundation of China under Grants No. 11905171 and No. 12047502. This work is also supported by the Natural Science Basic Research Plan in Shaanxi Province of China (Grant No. 2022JQ-025).

\end{document}